# QMe14S, A Comprehensive and Efficient Spectral Dataset for Small Organic Molecules

Mingzhi Yuan[1,#], Zihan Zou[1,#] and Wei Hu[*,1,2]

[1] School of Chemistry and Chemical Engineering, Qilu University of Technology (Shandong Academy of Science), Jinan 250353, China

[2] Key Laboratory of Precision and Intelligent Chemistry, School of Chemistry and Materials Science, University of Science and Technology of China, Hefei, Anhui 230026, China.

## ABSTRACT:

Developing machine learning protocols for molecular simulations requires comprehensive and efficient datasets. Here we introduce the QMe14S dataset, comprising 186,102 small organic molecules featuring 14 elements (H, B, C, N, O, F, Al, Si, P, S, Cl, As, Se, Br) and 47 functional groups. Using density functional theory at the B3LYP/TZVP level, we optimized the geometries and calculated properties including energy, atomic charge, atomic force, dipole moment, quadrupole moment, polarizability, octupole moment, first hyperpolarizability, and Hessian. At the same level, we obtained the harmonic IR, Raman and NMR spectra. Furthermore, we conducted ab initio molecular dynamics simulations to generate dynamic configurations and extract nonequilibrium properties, including energy, forces, and Hessians. By leveraging our E(3)-equivariant message-passing neural network (DetaNet), we demonstrated that models trained on QMe14S outperform those trained on the previously developed QM9S dataset in simulating molecular spectra. The QMe14S dataset thus serves as a comprehensive benchmark for molecular simulations, offering valuable insights into structure-property relationships.

## Background & Summary

Geometric Deep Learning[1,2] (GDL) is a subset of machine learning that extends traditional deep learning methods to data with geometric structures, such as graphs, manifolds, and point clouds. This makes GDL widely used in chemistry[3-5] and materials science[6-8], where molecules and materials can be represented as graphs. For example, Behler and Parrinello introduced a groundbreaking method that uses symmetry[9] functions to represent atomic distances and angles, allowing for the prediction of atomic properties. However, such feature engineering methods rely on manually constructed descriptors, which are limited by personal experience and permutation variations. In this context, message-passing neural networks (MPNNs)[10], which treat atoms and bonds (or inter-atomic distances) as a topological graph, propagate information between atoms and bonds to predict properties with permutation invariance or equivariance. Since 2017, many advanced MPNN models have been developed, such as SchNet[8,11], SpookyNet[12], EGNN[13], PaiNN[14], Nequip[15], Allegro[16], TeaNet[17], DimeNet[18], GEM[19], and TFNN[20]. Recently, we developed a deep equivariant tensor attention network[21] (DetaNet) that directly predicts accurate high-order tensorial molecular properties. Using DetaNet-predicted properties, we have also built modules to simulate IR, Raman, UV-Vis, and NMR spectra for organic molecules.

In addition to advanced machine learning models, datasets play a crucial role in determining the success of machine learning because the quality, quantity, and relevance of data directly impact model performance[22-24]. Without a sufficiently large and relevant dataset, models cannot learn effectively. Organic molecules exhibit vast diversity due to their molecular structures, bonding patterns, functional groups, isomerism, stereochemistry, incorporation of heteroatoms, chemical reactivity, and interactions with their environment[25-28].

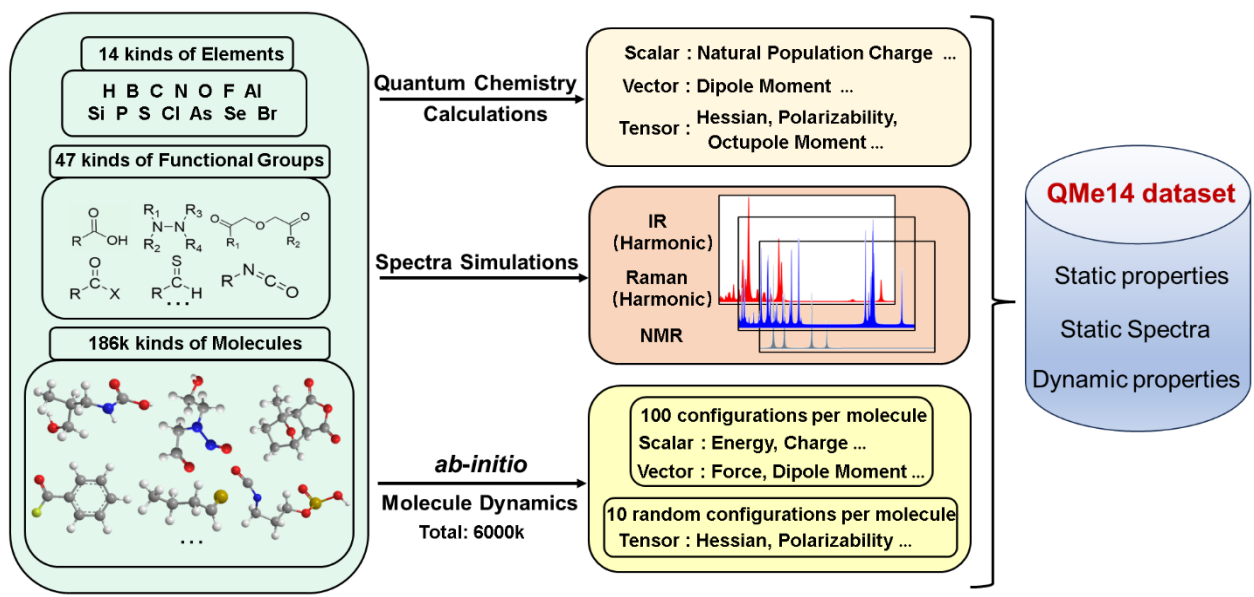

**Figure 1** Framework of constructing QMe14S dataset to cover 14 kinds of elements and 47 functional groups.

The generated database[29-32] (GDB) is an algorithmically created collection of organic molecules designed to explore the chemical space of small molecules. Many widely used datasets, such as QM7[31,33], QM8[32,34], QM9[32,35], and ANI-1[29,30,36], are subsets of GDB. While these datasets provide SMILES[37] (Simplified Molecular Input Line Entry System) representations for a large number of molecules within elemental constraints, they lack detailed geometric structures, molecular properties and spectroscopic data. To address this gap, we expanded the QM9 dataset by performing spectra (including IR, Raman, and UV-Vis) simulation at the B3LYP/TZVP level[38,39], constructing the QM9S dataset[21] (Data Citation 1). However, we found that the GDB generation algorithm does not comprehensively cover all possible chemotypes and functional groups, leading to biases in molecular representation. For instance, QM7 [31,33], QM8 [32,34], QM9[32,35], and ANI-1[29,30,36] include only 28, 27, 22, and 27 functional groups, respectively, and contain fewer than 10 types of elements. In contrast, datasets like QMugs[40] and PubChemQC[41] cover 42 and 45 functional groups, providing greater diversity for organic molecules. However, these datasets are so large that simulating the spectra using quantum chemistry or molecular dynamics becomes prohibitively expensive. Therefore, a key challenge remains in balancing the diversity of molecular species in the datasets with manageable computational costs.

In this work, we proposed QMe14S dataset by simulating IR, Raman, and NMR spectra for 186,102 organic molecules, encompassing 14 elements and 47 functional groups[42-44] (as illustrated in Fig. 1). We first analyzed the QM9S dataset, quantifying the occurrences of elements and functional groups. To enhance the dataset, we supplemented it with an additional 56,285 molecules from the PubChem database[45], ensuring that each element and functional group appears in over 500 instances. Quantum chemistry calculations were performed for all selected molecules to obtain molecular properties and corresponding spectra. Furthermore, we conducted ab initio molecular dynamics simulations to capture the dynamic properties of nonequilibrium conformations. The proposed QMe14S dataset offers a more extensive and efficient molecular spectral resource, which can serve as a benchmark for molecular simulations and structure retrieval.

## Methods

**Data collection and preprocessing.** Based on the statistical analysis on the QM9S dataset, we firstly supplemented 56,285 molecules from PubChem website using sub-structure search function of RDKit (http://www.rdkit.org) to ensure every element and functional group appears in over 500 instances. The added functional groups contain carboxyl, hydrazine, and hydrazone groups, while the added elements are Cl, S, P, Br, B, Si, As, Se and Al. In addition, we also filtered several 'token' that are commonly exist in SMILES[37] and represent different chemical spaces, such as double bond ('='), triple bond ('#'), branch ('(' and ')'), cis-trans isomerism ('/' and '\'), and chiral center ('@' and '@@'). During filtering, we selected the molecules as small as possible to reduce the QC computational cost. In this way, we added the molecular diversity in terms of compound type and elemental composition. We saturated all SMILES in QMe14S with hydrogen atoms using RDKit, and subsequently generated into the initial Cartesian coordinates.

**Quantum mechanical calculation details.** We then performed geometry optimization and frequency analysis on the 56,285 molecules added to QMe14S, using the Gaussian 16 software package[46] at the B3LYP/TZVP level[38-39]. From the equilibrium configurations, we obtained scalar, vectorial, and tensorial properties, as well as IR, Raman, and NMR spectra. Combining the QM results of these 56,285 molecules with the QM9S dataset yielded the QMe14S dataset, a comprehensive collection of properties and spectra for 186,102 molecules.

We conducted ab initio molecular dynamics simulations at the B3LYP/TZVP level using the atom-centered density matrix propagation[47] (ADMP) method in Gaussian 16 to capture dynamic proper-ties. The step size and total simulation time were set to 1 femtosecond and 100 femtoseconds, respectively, resulting in 6 million molecular dynamic configurations with dynamic properties such as energies, atomic forces, and dipole moments. From these, we randomly selected 10 conformations out of every 100 for each molecule to calculate polarizabilities and Hessian matrices at the same level of theory.

**Training details.** We tested and validated the QMe14S dataset using our developed deep equivariant tensor attention network[21] (DetaNet) considering its ability of predicting high-order tensorial properties. We selected the dynamic properties such as energy and atomic forces for ML training because thy play important roles in molecular dynamic simulations. All molecules were divided into training, validation and test sets with the ratio of 90%, 5% and 5%, respectively. Here we should keep the 47 functional groups appear evenly in the three set to keep the diversity of any set. Moreover, the 100 dynamic con-formations of each molecule should be packaged into the same set (training, validation or test) to enhance the DetaNet's transferability.

To simulate the IR and Raman spectra, we trained DetaNet to predict the Hessian matrix, dipole moment derivatives, and polarizability derivatives. By diagonalizing the Hessian matrices, we obtained the vibrational frequencies and normal modes. Combining the dipole moment and polarizability derivatives with the normal modes, we calculated the IR and Raman intensities for each mode. Finally, we generated the IR and Raman spectra using Lorentzian broadening with a half-width of 15 and 10 cm$^{-1}$, respectively. Additionally, we trained DetaNet to predict the chemical shift and isotropic values of the magnetic shielding tensor for NMR spectrum simulations.

## Data Records

The QMe14S dataset is hosted in the Figshare data repository. It includes all associated molecular structures, along with their corresponding properties and spectra, which are publicly accessible. A README file provides detailed instructions, and an example usage is demonstrated in the read_dataset.py script.

**File format**

All properties of each molecular structure is stored in torch.geometric.data.Data[48] in HDF5 format[49]. Table 1 lists the symbol, dimension and unit of every property. The OPT_186102.h5 file contains the optimized geometries and corresponding static properties of the 186,102 molecules. The MD_x.h5 files store the dynamic configurations and corresponding properties (energy, force and dipole moment) from the MD simulations, each single file containing 1 million configurations for 10 thousands molecules. The QMe14S_single_point.h5 file stores the 10 configurations randomly selected from 100 dynamic configurations and the corresponding dynamic properties (charge, quadrupole, octapole, polarizability and electronic spatial extent). The Hessian matrix of the equilibrium (optimized) and nonequilibrium (dynamic) geometries are stored into Hessian_opt.h5 and Hessian_single_point.h5. The IR_broaden.csv and Raman_broaden.csv file store the Lorentz broadened IR and Raman spectra, while NMR.h5 file stores the magnetic shielding tensors.

**Table.1** The properties covered by QMe14S, as well as the corresponding symbol, dimension, unit in storage files.

| No. | Property | Symbol | Dimension | Unit | Total Numbers |
|---|---|---|---|---|---|
| 1 | SMILES | — | — | — | 186,102 |
| 2 | Atomic numbers | Z | N | — | 186,102 |
| 3 | Atomic positions (coordinates) | R | 3N | Å | 6,093,102 |
| 4 | Atomization Energy | E | 1 | eV | 5,907,000 |
| 5 | Atomic Force | F | 3N | eV/Å | 5,907,000 |
| 6 | APT Charges | δ | N | e | 539,737 |
| 7 | Mulliken Charges | δ | N | e | 539,737 |
| 8 | Natural Population Charge | δ | N | e | 186,102 |
| 9 | Electronic spatial extent | $R^2$ | 1 | $a_0^2$ | 539,737 |
| 10 | Dipole Moment | μ | 3 | D | 6,093,102 |
| 11 | Polarizability | α | 9 | $a_0^3$ | 725,839 |
| 12 | First Hyperpolarizability | β | 27 | a.u. | 186,102 |
| 13 | Quadrupole Moment | $Q_{ij}$ | 9 | D·Å | 725,839 |
| 14 | Octupole Moment | $\Omega_{ijk}$ | 27 | D·Å$^2$ | 725,839 |
| 15 | Hessian | H | $9N^2$ | eV·Å$^{-2}$ | 725,839 |
| 16 | Dipole Derivatives | $-\nabla_{Ri}\mu$ | 9N | D·Å$^{-1}$ | 725,839 |
| 17 | Polarizability Derivatives | $-\nabla_{Ri}\alpha$ | 18N | $a_0^3$·Å$^{-1}$ | 186,102 |
| 18 | Shielding Tensor | σ | 9N | ppm | 59,260 |
| 19 | IR_broaden | — | 3500 | — | 725,839 |
| 20 | Raman_broaden | — | 3500 | — | 186,102 |

## Technical Validation

The diversity of elements and functional groups represent the chemical space of the dataset. As shown in Fig. 2a, QMe14S includes 14 elements: H, B, C, N, O, F, Al, Si, P, S, Cl, As, Se, and Br. On the other hand, QM7, QM8, QM9, QMugs, PubChemQC and ANI-1 contains 5, 5, 5, 10, 36 and 5 types of elements. However, QMugs and PubchemQC contain 665,911 and 3,982,436 molecules, respectively, making full quantum chemistry calculations on these datasets unfeasible. The standard deviation of the occurrences of each element in QMe14S is approximately 266,537, compared to 4,770,545 and 7,120,639 in QMugs and PubChemQC, respectively. This relatively uniform distribution of elements in QMe14S significantly reduces computational cost. Beyond elements, QMe14S also includes 47 types of functional groups (Fig. 2b). The standard deviation for functional group occurrences in QMe14S is 28,954, compared to 190,165 and 821,707 for QMugs and PubChemQC. This relatively uniform distribution of functional groups further balances high chemical space coverage with improved efficiency in subsequent quantum chemistry computations.

Beyond elements and functional groups, the diversity of spatial structures also defines the chemical space. For example, 1D (rod-shaped), 2D (disc-shaped), and 3D (near-spherical) molecules exhibit distinct physical and chemical properties. We used the principal moments of inertia[50] (PMI) diagram to illustrate the diversity of molecular shapes within QMe14S, with color-coded atomization energies from density-functional theory representing molecule stability. As shown in Fig. 2c, QMe14S encompasses various molecular shapes, although near-spherical molecules are less common. We also conducted principal component analysis[51] (PCA) on the QMe14S, QM7, QM8, and ANI-1 datasets. The SMILES representations were converted to Morgan fingerprints using RDKit and scaled collectively to visualize the chemical space distribution. As shown in Fig. 2d-f, the chemical space of QMe14S is significantly broader than that of QM7, QM8, and ANI-1.

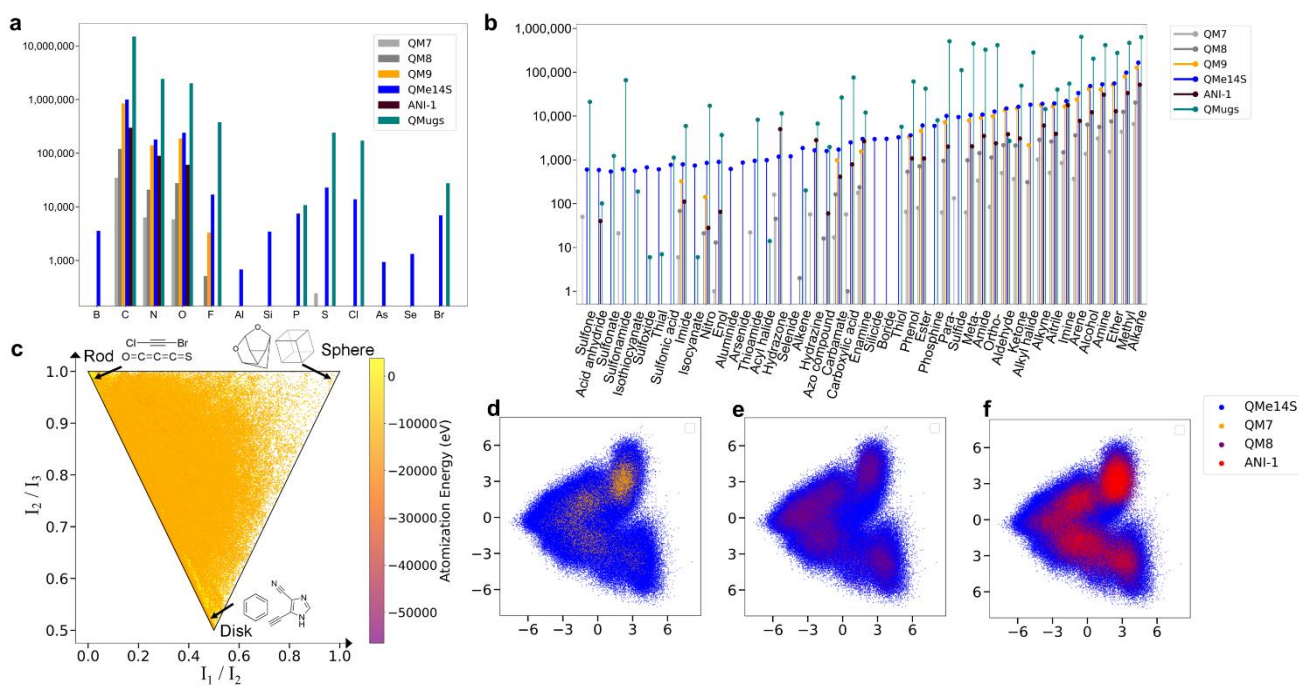

**Figure 2** (**a, b**) Diversity comparison of the elements and functional groups for the QMe14S with QM7, QM8, QM9, ANI-1 and QMugs. (**c**) Principal-moments-of-inertia plot for molecules in the QMe14S dataset. The x and y axes are normalized moments of inertia ($I_1/I_2$ and $I_2/I_3$), $I_x$=x-th the smallest principal moment of inertia. (**d-f**) Principal component analysis of QMe14S versus comparably sized QM7, QM8 and QM9 datasets.

The comparison of the QMe14S with QM7[31,33], QM8[32,34], QM9[32,35], QMugs[40], PubchemQC[41], QM7-X[31,52], ANI-1[29,30,36] are listed in Table 2. We can see that QM7, QM8, QM9, QMugs, PubchemQC, QM7-X, ANI-1, QMe14S dataset are obtained at PBE0/ Tier2, RI-CC2/def2TZVP, B3LYP/6-31G(2df, p), GFN2-xTB, B3LYP/6-31G(d), PBE0+MBD, ωB97-X/6-31G(d), B3LYP/TZVP levels, respectively. From Table 2, we can see that QM7, QM8, QM9, and PubChemQC contain only the static properties of the equivalent configuration, while QM7-X, QMugs, ANI-1, and QMe14S also reported the dynamic properties of the nonequilibrium structures. The atomic forces of the dynamic configurations are only contained in QM7-X and QMe14S. The datasets QM9, QMugs, PubChemQC, QM7-X, and QMe14S provide dipole moments, with QM9, QMugs, QM7-X, and QMe14S also including polarizabilities, though QMugs' values are calculated using semi-empirical methods. Quadrupole moments are found in QM7-X, QMe14S, and QMugs (semi-empirical), while higher-order tensors like the first hyperpolarizability, octupole moments, Hessian matrix, and property derivatives are uniquely included in QMe14S. The QM8 and PubChemQC datasets provide UV-

Vis spectral data, with QM8 also including electronic spectra. IR spectra are available in QMugs and QMe14S, while Raman spectra are unique to QMe14S. Additionally, QMe14S includes NMR spectra for 60k molecules selected from PubChem.

Table.2 Comparison of QMe14S with QM7, QM8, QM9, QMugs, PubchemQC, QM7-X, ANI-1 dataset in coverage of elements, molecular size, computational method, molecular properties and spectroscopic data.

| | | QM7 | QM8 | QM9 | QMugs | PubChemQC | QM7-X | ANI-1 | QMe14S |
|---|---|---|---|---|---|---|---|---|---|
| Unique Compounds | | 7,165 | 21,800 | 133,885 | 665,911 | 3,982,436 | 7,211 | 57,462 | 186,102 |
| Source | | GDB-13 | GDB-17 | GDB-17 | ChEMBL | PubChem | GDB-13 | GDB-11 | PubChem &GDB-17 |
| Number of Element Species | | 5 | 5 | 5 | 10 | 36 | 6 | 5 | 14 |
| Heavy Atoms max | | 7 | 8 | 9 | 100 | 51 | 7 | 8 | 32 |
| Method | | PBE0/ tier2 | LR-TDPBE0/ def2-SVP +RI-CC2/ def2-TZVP | B3LYP/ 6-31G(2df,p) | GFN2-xTB+ ωB97X-D/ def2-SVP | B3LYP/ 6-31G(d) | PBE0+ MBD | ωB97X/ 6-31G(d) | B3LYP/ TZVP |
| Multiple Configurations Sampling | | × | × | × | √ | × | √ | √ | √ |
| Property | Excited State Energy | × | √ | × | × | √ | × | × | × |
| | Atomization Energy | √ | × | √ | √ | √ | √ | √ | √ |
| | Atomic Force | × | × | × | × | × | √ | × | √ |
| | Mulliken charges | × | × | √ | √ | √ | × | × | √ |
| | APT charges | × | × | × | × | × | × | × | √ |
| | Natural Population Charge | × | × | × | × | × | × | × | √ |
| | Dipole Moment | × | × | √ | √ | √ | √ | × | √ |
| | Polarizability | × | × | √ | √(GFN2) | × | √ | × | √ |
| | First Hyperpolarizability | × | × | × | × | × | × | × | √ |
| | Quadrupole Moment | × | × | × | √(GFN2) | × | √ | × | √ |
| | Octupole Moment | × | × | × | × | × | × | × | √ |
| | Hessian | × | × | × | × | × | × | × | √ |
| | Dipole Derivatives | × | × | × | × | × | × | × | √ |
| | Polarizability Derivatives | × | × | × | × | × | × | × | √ |
| Spectra | IR | × | × | × | √ | × | × | × | √ |
| | Raman | × | × | × | × | × | × | × | √ |
| | NMR | × | × | × | × | × | × | × | √ |
| | UV-Vis | × | √ | × | × | √ | × | × | × |
| | Electronic spectra | × | √ | × | × | × | × | × | × |

**ML-Predicted Properties.** We firstly trained the DetaNet on the QMe14S dataset to predict the molecular scalar, vectorial and tensorial properties. Prediction accuracy was evaluated against reference DFT data (B3LYP/TZVP level) using mean absolute error (MAE), root mean square error (RMSE)[53], and the coefficient of determination[54] ($R^2$). As shown in Fig. 3a, b, dynamic molecular properties like energies and forces were predicted with accuracies of 99.999% and 99.858%, respectively. On the other hand, the static molecular properties including the natural charge analysis, dipole moments, and polarizability can also be well predicted with accuracies over 99.9%, as shown in Fig. 3c-e. We also accessed the ML accuracy in predicting the derivative of several properties that is related to the spectra simulations. As shown in Fig. 3(f, g), the average absolute error of the dipole moment derivatives, polarizability derivative and Hessian matrix are 0.0395 D·Å$^{-1}$, 0.467 and 0.107 eV·Å$^{-2}$, respectively, with accuracy 99.630%, 97.668% and 99.896%.

**ML-Predicted Spectra.** Using ML-predicted molecular properties, we evaluate DetaNet's accuracy in predicting molecular vibrational spectra. Here we examined two DetaNet models, trained on QM9S and QMe14S datasets (referred to as DetaNet- QM9S and DetaNet-QMe14S), to compare with the DFT-predicted spectra. From Fig. 4a–f, we can observe that DetaNet-QM9S performs poorly in simulating molecular IR and Raman spectra. For example, the N-N bond stretching vibrational mode of the morpholin-4-amine, azepan-1-amine, and octane-hydrazide molecules that is predicted at around 842 $cm^{-1}$ in the DFT calculations, shows red-shift and enhanced intensities in DetaNet-QM9S model. Furthermore, we observed the red-shifted N-H bond bending vibrational mode (1676 $cm^{-1}$) and the misestimated IR intensities of the N-H bond stretching vibrational peaks (3417 $cm^{-1}$) for morpholin-4-amine, azepan-1-amine, and octanehydrazide molecules (shown in Fig. 4a-c). On the other hand, the DetaNet-QM9S simulated Raman spectra shows red-shifted N-H stretching vibrational peaks (3452 $cm^{-1}$) for 3-hydrazineylpropanoic acid and ethoxycarbamic acid compared to the DFT simulations, as shown in Fig. d-e. For the 3-phenyl-1-vinyltriaz-1-ene molecule, the DetaNet-QM9S underestimated the Raman intensity of the N=N stretching (1456 $cm^{-1}$) and N-N stretching (1197 $cm^{-1}$) modes, while overestimated the intensities of the phenyl ring C-H (3059-3103 $cm^{-1}$), as shown in Fig. 4(f).

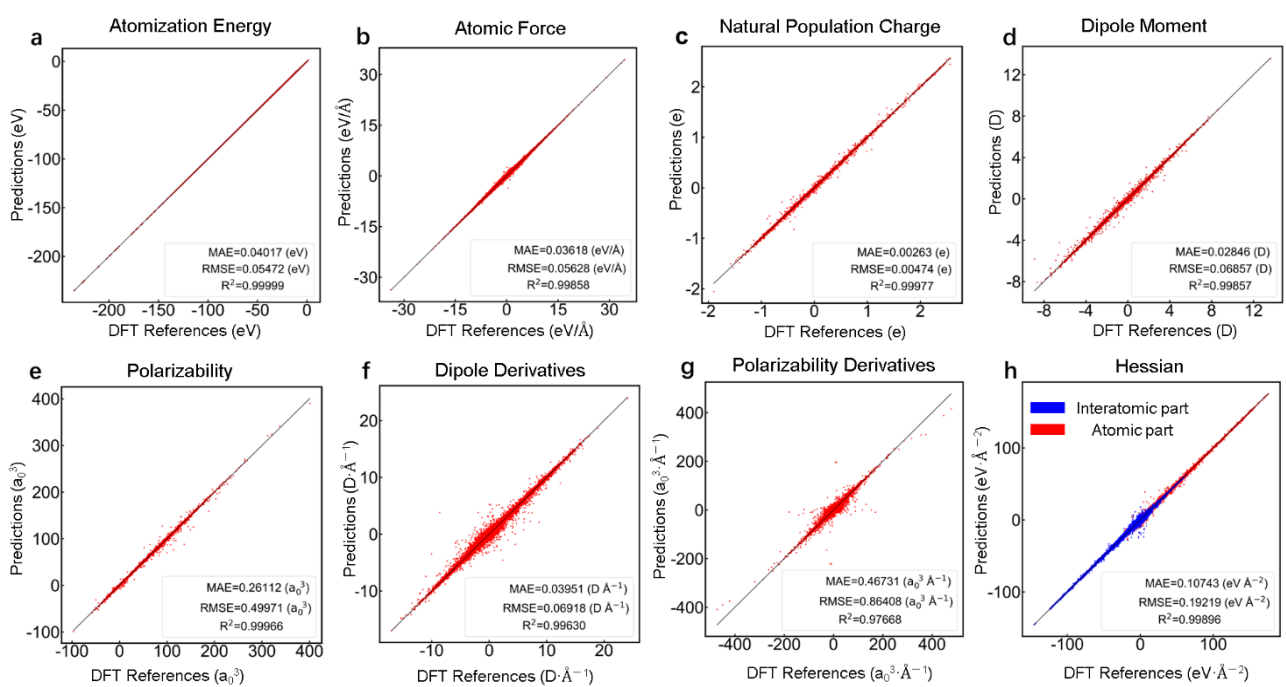

**Figure 3** Error distributions and regression plots of DetaNet's predictions training on QMe14S for eight properties. (**a, b**) The dynamic properties of energy and atomic force. (**c-h**) The static properties of natural population charge, electric dipole moment, polarizability, first hyperpolarizability, electric quadrupole moment, and electric octupole moment. The MAE, RMSE and R2 represent the mean absolute errors, the root mean square errors and the coefficients of determination.

We used the root mean square error[53] (RMSE), cosine similarity[55], Pearson and Spearman correlation coefficients[56] to assess the spectral similarity between the DetaNet-predicted and DFT-predicted vibrational spectra. We found the average values of the RMSE, cosine, Pearson, and Spearman coefficients between the DetaNet-QMe14S predicted and DFT-predicted IR spectra for the 7038 molecules containing only HCNOF elements in the test set were about 0.0348, 94.63%, 93.78% and 99.37%, respectively. These values of the Raman spectra were 0.0329, 92.34%, 91.58% and 99.11%, respectively. In case of DetaNet-QM9S predictions, these values are 0.0348, 93.84%, 92.96%, 99.25% for IR spectra, and 0.0328, 91.46%, 90.67%, 99.01% for Raman spectra.

To simulate the NMR spectra, we predicted the isotropic magnetic shielding tensor of C and H atoms and the corresponding chemical shift values. We can see from Fig. 4(g-i) that the DateNet-QMe14S predicted 13C NMR spectra of hexylcarbamic acid and (3-hydroxy-2-methylpropyl) carbamic acid molecules and the 1H NMR spectra of N,N-dibutylnitrous amide molecules, are almost the same as the DFT calculations, demonstrating the comprehensiveness and efficiency of the QMe14S spectral dataset.

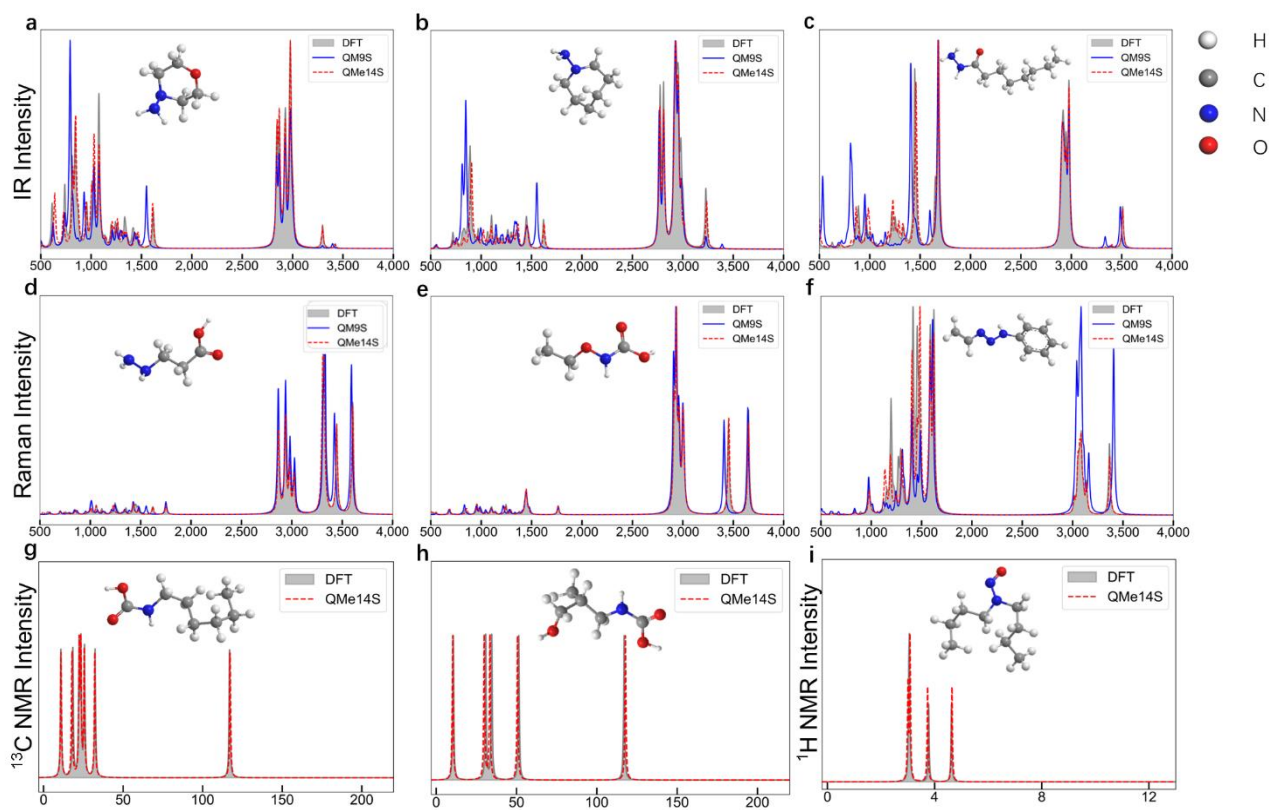

**Figure 4** Compared molecular spectra obtained from the DetaNet model trained on the QM9S and QMe14S dataset with DFT simulations. (**a-c**): IR spectra of morpholin-4-amine, azepan-1-amine and octanehydrazide molecule. (**d-f**) Raman spectra of 3-hydrazineylpropanoic acid, ethoxycarbamic acid and 3-phenyl-1-vinyltriaz-1-ene molecule. (**g, h**) 13C NMR spectra of hexylcarbamic acid and (3-hydroxy-2-methylpropyl) carbamic acid molecule. **i** 1H NMR spectra of N,N-dibutylnitrous amide molecule.